# Determination of the sheet resistance of an infinite thin plate with five point contacts located at arbitrary positions

Krzysztof R. Szymański, Piotr A. Zaleski
Faculty of Physics, University of Białystok, K. Ciołkowskiego Street 1L,
15–245 Białystok, Poland

**Abstract**

In this paper, a five-probe method of sheet resistance measurement that is independent of probe positions is reported. The method is strict for an infinite homogeneous plane. It has potential applications as a sheet resistance standard based on planar molecular layers. The method can be used to measure the sheet resistance of layers covering objects with a spherical topology, particularly on micro- and nanometric scales, where it is difficult to control probe positioning.

## 1. Introduction

Conducting materials with thicknesses that are much less than their planar dimensions exhibit a natural ratio of electrical resistivity to thickness, called sheet resistance. The principle of sheet resistance is being revisited owing to recent technologies providing access to naturally planar materials, such as graphene and other atomic layers, for which thickness does not, whereas sheet resistance does, have a precise meaning. Well-established four-probe point contact methods of sheet resistance measurement are being continuously improved [2,3,4]. The four-probe point contact methods available are described in a recent review by Miccoli et al. [1]. The most important are the in-line four probe, square four probe, and van der Pauw methods [5]. The first two techniques are essentially local, with their precision depending on the precision of the contact arrangement and the precision of the resistance measurement itself. The third is a global technique in which four probes are located on the sample edge. Briefly, in the van der Pauw method, the function $f(\rho, r_1, r_2)$ is defined

$$f(\rho, r_1, r_2) = \exp\frac{-\pi r_1}{\rho} + \exp\frac{-\pi r_2}{\rho}, \tag{1}$$

where $r_1$ and $r_2$ are the usual two four-probe resistances for four contacts located arbitrarily on the edge of a thin, homogeneous sample [5]. The function (1) has a property, that

$$f(\rho_0, r_1, r_2) = 1, \tag{2}$$

where $\rho_0$ is the sheet resistance.

In this paper, we propose a new method of sheet resistance measurement in which contacts are located far from sample edges and the sheet resistance measurements obtained are independent of contact site locations.

## 2. Five-point method

Let us consider four point contacts at arbitrary positions $a$, $b$, $c$, and $d$ on a homogeneous, isotropic, infinite, thin conducting plane. We define four-probe resistance in a standard way as $r_{abcd} = V_{cd}/j_{ab}$, where current $j_{ab}$ enters the sample at contact $a$ and leaves at contact $b$, and the potential $V_{cd}$ is measured between contacts $c$ and $d$ (Fig. 1a).

We construct the function $f(\rho, r_1, r_2, r_3, r_4, r_5)$ wherein the unit of parameter $\rho$ is the same as the sheet resistance unit:

$$f(\rho, r_1, r_2, r_3, r_4, r_5) = \sum_{i=1}^{5}(x_i - x_i x_{i+2} - x_i x_{i+1} x_{i+3} + x_i x_{i+2}^2 x_{i+4}) - x_1 x_2 x_3 x_4 x_5. \quad (3)$$

In eq. (3), we use a cyclic index $x_{i+5} = x_i$ to define $x_i$ for indices outside the range of 1,…5. The parameters $x_i$ are introduced to shorten the exponential notation, such that

$$x_i = \exp\frac{-4\pi r_i}{\rho}, i = 1,\ldots 5. \quad (4)$$

The standard four-probe resistances $r_1, r_2, r_3, r_4, r_5$ are

$$r_1 = r_{abcd}, r_2 = r_{bcde}, r_3 = r_{cdea}, r_4 = r_{deab}, r_5 = r_{eabc}, \quad (5)$$

defined for all subsets of 4 contact points among five points *abcde* chosen on an infinite plane (Fig. 1). The function defined in eq. (3) has an important property, that

$$f(\rho_0, r_1, r_2, r_3, r_4, r_5) = 1, \quad (6)$$

where $\rho_0$ is the sheet resistance of the infinite plane. Eq. (6) is analogous to eq. (2). The main difference between them is that in the van der Pauw method represented in eq. (2), the four probes are located at arbitrary positions along the edge, while in the five-point method represented in eq. (6), the probes need to be located far from the edge.

The derivation of eq. (6) is presented in Appendix 1. Irrespective of this derivation, one may introduce eq. (3) directly into the left side of eq. (6) and show that eq. (6) yields a value of one, if the proper values of $r_i$ are used (see eq. A2 in Appendix 1). Because eq. (6) is valid for arbitrary positions of the contacts *abcde* on an infinite plane, it can be used for measuring sheet resistance in a manner analogous to the van der Pauw method. By measuring the five resistances $r_i$, one can obtain numerical solutions to eq. (6) in which there is only one unknown variable, namely $\rho$, and the solution for eq. (6) is the sheet resistance $\rho_0$. The method is similar to well-known four-probe methods in which the positions of four contacts are required to fulfill a set of rules. However, in contrast to these methods, the five-point method enables sheet resistance to be measured independently of the locations of the contact positions.

Let us briefly discuss some properties of eq. (6). Formally, the variables $r_1, r_2, r_3, r_4, r_5$ in eq. (3) are not truly independent, because the $r_i$ resistances need to be defined correctly according to a predetermined order of contacts *abcde*, as shown in eqs. (5). The arguments in function $g(\rho, a, b, c, d, e)$, which is introduced in Appendix 2, are independent variables. One may always use the relationship $f(\rho, r_1, r_2, r_3, r_4, r_5) - 1 = g(\rho, a, b, c, d, e)$ in eq (3). These remarks apply also to the function in eq. (1), which is used in the van der Pauw method.

It is clear from the construction of eq. (3) that eq. (6) is valid for any permutation of contacts. However, when one solves eq. (6) to obtain $\rho_0$, the precision of the answer depends on the order of the *abcde* contacts and thus on the permutation of the $r_i$ arguments in the function in eq. (3). Given the reasonable assumption that the uncertainties of all measured values of $r_i$ are the same, equal to $\delta r$, the precision of the sheet resistance value $\delta\rho$ extracted from eq. (6) can be estimated as:

$$(\delta\rho_0)^2 = \left(\frac{\partial f}{\partial \rho}\right)^{-2} \sum_{i=1}^{5} \left(\frac{\partial f}{\partial r_i}\right)^2 (\delta r)^2 \stackrel{\text{def}}{=} u^2(\delta r)^2. \quad (7)$$

Below, we discuss some general properties of a homogeneous plane with five contacts. Let us first consider an arbitrary circuit in which there are resistances measured at five contacts, not necessarily in a homogeneous plane. For each set of four contacts, there are 24 values of resistance.

These 24 measurements can be grouped into three groups of 8. In each group four contact choices result in the same value of resistance, and four others result in resistances of the opposite sign. We thus have, in principle, only three different resistance values.

For a set of five arbitrary contact locations, there are five possible choices of four elements. Therefore, for five contacts, we can in principle obtain $2 \cdot 3 \cdot 5 = 30$ different resistance values. Because of the reciprocity theorem, $r_{abcd} = r_{cdab}, r_{abcd} + r_{adbc} = r_{acbd}$ [5], and we have a sign constraint $r_{abcd} = -r_{bacd}$. Further, along with the requisite constraint for the five probe system $r_{abcd} + r_{abde} = r_{abce}$, these relations limit the number of independent resistances to five. Moreover, on a homogeneous plane, with application of the constraint given by (6), the number of independent parameters is further reduced to four, and they are given explicitly by $\mu_1, \nu_1, \mu_2, \nu_2$ in eqs. (A3)–(A7) of Appendix 1.

For a given order of contacts *abcde*, there are five circular permutations generated by moving the final element to the front (i.e. *abcde* becomes *eabcd*). The same principle applies for the reversed order (i.e. *edcba* becomes *aedcb*). For all ten permutations, the uncertainty parameter $u$ in eq. (7) has the same value. Therefore, one expects to have $5!/10 = 12$ nonequivalent permutations, and thus 12 different values of $u$. Some examples are listed in Table 1 and shown in Fig. 2. The physical meaning of nonequivalent permutations is clear, with each one corresponding to one of the loops shown in Fig. 1*c*.

It is of interest to see eq. (6) solutions for different nonequivalent permutations of contacts *abcde*. When we calculate resistances $r_i$ (5) for the contact arrangements shown in Fig. 2 on a plane with sheet resistance $\rho_0$, and determine values of the function $f$ for nonequivalent permutations of contacts *abcde* (Fig. 3), we find that the number of non-overlapping branches in Fig. 3*a-f* does not coincide with the number of different $u$ values in Table 1. In all cases, the function $f$ is equal to 1 when $\rho = \rho_0$, as expected from eq. (6). In some cases, nonphysical solutions appear for $\rho < \rho_0$ (see $\rho \approx 0.4\rho_0$, $\rho \approx 0.92\rho_0$, and $\rho \approx 0.75\rho_0$ for Fig. 3*b*, *e*, and *f*, respectively). We find that all solutions for different permutations of *abcde*, represented by curves in Fig. 3, cross the horizontal red line representing $\rho = \rho_0$ at two points. This can be explained as follows.

Let us take the function in eq. (1) for contact order *abcde*, with $f(\rho, r_1, r_2, r_3, r_4, r_5)$ as defined in eq. (3). When we take another contact order, for example *a'b'c'd'e'* = *bacde*, define $r'_1 = r_{bacd}$, $r'_2 = r_{acde}$, $r'_3 = r_{cdeb}$, $r'_4 = r_{deba}$, and $r'_5 = r_{ebac}$, and consider $f(\rho, r'_1, r'_2, r'_3, r'_4, r'_5)$ as defined in eq. (3), it can be shown that

$$\ln\frac{f(\rho, r_1, r_2, r_3, r_4, r_5) - 1}{f(\rho, r'_1, r'_2, r'_3, r'_4, r'_5) - 1} = \frac{\lambda}{\rho}, \tag{8}$$

where $\lambda$ depends only on the contact positions and the chosen permutation *bacde*. Thus, if the function $f(\rho, r'_1, r'_2, r'_3, r'_4, r'_5) - 1$ reaches zero for $\rho = \rho_0$, then function $f(\rho, r_1, r_2, r_3, r_4, r_5) - 1$ reaches zero as well for $\rho = \rho_0$. This result is valid for $\rho$ having a sheet resistance of $\rho_0$ and also for non-physical solutions. In Appendix 2, we prove that the property represented in eq. (8) is valid for any two permutations *abcde* and *a'b'c'd'e'*.

Origin of the second solution results in nonlinearity of eq. (3) in parameter $\rho$. To show it, let us introduce polynomial of $x_i$ using the definition (3): $P(x_1, x_2, x_3, x_4, x_5) \overset{\text{def}}{=} f(\rho, r_1, r_2, r_3, r_4, r_5) - 1$. It is clear from (4) that small change of all resistances $r_i \to r'_i = (1 + \varepsilon)r_i$ results in the change of $x_i \to x'_i = x_i^{1+\varepsilon}$, and the expansion of polynomial $P(x'_1, x'_2, x'_3, x'_4, x'_5)$ up to the second order in $\varepsilon$ yields

$$P(x'_1, x'_2, x'_3, x'_4, x'_5) = P(x_1, x_2, x_3, x_4, x_5) + c_1\varepsilon + c_2\varepsilon^2 + O(\varepsilon^3), \tag{9}$$

where

$$c_1 = \sum_{i=1}^{5}(x_i \ln x_i - x_i x_{i+2} \ln x_i x_{i+2} - x_i x_{i+1} x_{i+3} \ln x_i x_{i+1} x_{i+3} \tag{10}$$
$$+ x_i x_{i+2}^2 x_{i+4} \ln x_i x_{i+2}^2 x_{i+4}) - x_1 x_2 x_3 x_4 x_5 \ln x_1 x_2 x_3 x_4 x_5,$$

$$c_2 = \frac{1}{2}\sum_{i=1}^{5}(x_i(1 - x_{i+2} - x_{i+4})\ln^2 x_i - 2x_i x_{i+2}\ln x_i \ln x_{i+2} - x_i x_{i+1} x_{i+3} \ln^2 x_i x_{i+1} x_{i+3} \tag{11}$$
$$+ x_i x_{i+2}^2 x_{i+4} \ln^2 x_i x_{i+2}^2 x_{i+4}) - \frac{1}{2} x_1 x_2 x_3 x_4 x_5 \ln^2 x_1 x_2 x_3 x_4 x_5.$$

As in (3), we have used in (10) and (11) a cyclic index $x_{i+5} = x_i$ to define $x_i$ for indices outside the range of 1,…5. From (6) we have $P(x_1, x_2, x_3, x_4, x_5) = 0$ for $\rho = \rho_0$, and if $\varepsilon = -c_1/c_2$ eq. (9) is equal to zero. Thus for set of parameters $x'_1, x'_2, x'_3, x'_4, x'_5$ function $f(\rho, r'_1, r'_2, r'_3, r'_4, r'_5) = 1$ for $\rho$ equal to some value of $\rho'$. From construction (3-5) we have $f(\rho, r_1, r_2, r_3, r_4, r_5) = f(k\rho, kr_1, kr_2, kr_3, kr_4, kr_5)$, for any $k > 0$. Therefore ratio $\rho'/\rho_0$ has the same value as the ratio $r'_i/r_i$ and

$$\rho' = \left(1 - \frac{c_1}{c_2}\right)\rho_0. \tag{12}$$

Numerical estimations from (12) are in agreement with second solution of (6) shown in Fig. 3. In case of arrangement shown in Fig. 3 *d* $\varepsilon=0$ while for arrangements shown in Fig 3 *e, f* conditions correspond to regime of small $\varepsilon$.

We are interested in knowing the contact arrangement that provides the most precise experimental result, which occurs when the value of $u$ has a minimal value. By performing numerical simulations with randomly generated contact locations, we find that the value of $u$ in eq. (7) can be as small as 6.8, though the absolute lower limit is not known. Also, the positions of the contacts for this lower limit are not unique. This ambiguity exists because currents and potentials are invariant under conformal mapping, while Euclidean distances not. Thus, an infinite number of contact configurations could yield the same values of resistances $r_i$, or the same uncertainty parameter $u$. However, not all of these configurations are equivalent. Configurations in which two contacts are close relative to the distances between other contacts are unwanted. To quantify the problem, let us introduce the packing ratio $p$ for a finite set of points located on the plane as

$$p = \frac{\text{Min}\{d_{ij}\}}{\text{Max}\{d_{ij}\}}, \quad i \neq j, \tag{13}$$

where $d_{ij}$ is the Euclidean distance between a pair of contacts $i$ and $j$. We are interested in adopting contact configurations for which $p$ is as large as possible. The theory of complex numbers posits that a map of the Riemann sphere onto itself is conformal if and only if it is a Möbius transform. The same principle applies to a complex plane. Thus, without loss of generality, we will consider Möbius transforms that maximize eq. (13). During our computer simulations searching for a minimal value of uncertainty $u$, we have also calculated the packing ratios $p$ (Fig. 4, results shown as black points). The arrangements that yield the smallest value of $u$ can be found and subjected to a Möbius transform to maximize $p$ (green point *f* in Fig. 4 and in Fig. 2*f*). One may thus expect that the arrangement shown in Fig. 2*f* optimizes the experimental precision that can be obtained.

Regarding specific contact arrangements, we find, quite remarkably, that for four contacts located on a circle or on a line, the parameter $u$ in eq. (7) is independent of the location of the fifth contact (proof in Appendix 3). Examples of these arrangements are shown in Fig. 2*a-e*. The lowest value of $u$ for four contacts located on the circle is about 7.66 (blue line in Fig. 4). Because $u$ does not depend on the location of the fifth contact, we use these degrees of freedom for maximizing $p$ by way of a Möbius transform and achieve a value of 0.554 (Fig. 2e and Table 1).

In the case of contacts *abcde* being located at the edges of a regular pentagon ($x_i = x$ for $i = 1,...5$), the identity (6) reduces to:
$$5x - 5x^2 - 5x^3 + 5x^4 - x^5 = 1, \tag{14}$$
and is fulfilled for $x = \tau + 1 \approx 2.618$, where the parameter $x$ corresponds to the case that all resistances $r_i = r$, with
$$r = \frac{\rho}{4\pi}\ln(\tau + 1), \tag{15}$$
$\tau = (1 + \sqrt{5})/2$ is the golden ratio. The possible distinct values of four-point resistance are $\pm r$ and $\pm 2r$. In particular, if the points *abcde* are ordered clockwise or anti-clockwise around a pentagon perimeter, $r_{abcd} = r_{acbd} = r$ and $r_{adcb} = r_{abce} = 2r$. For a regular pentagon, the packing ratio $p$ achieves its highest possible value at $\tau - 1 \approx 0.618$. This maximal value, represented by the vertical red line in Fig. 4, is also a limit for the values shown by the black points in Fig. 4. Computer simulations designed to reveal the conditions for which high $u$ values are achieved suggest that there is no upper limit for $u$ when excessive closeness of contacts in two pairs results in a small value of $p$.

## 3. Effect of finite size

The plane represented in eq. (6) is infinite, while measurements are usually performed on finite thin plate. Employing elementary calculations based on the concept of images in electrostatics, the four-probe resistance obtained for contacts on a finite disk with radius $R$ is
$$r_{abcd} = \frac{\rho}{2\pi}\left(\ln\frac{|ac|\,|bd|}{|ad|\,|bc|} + \ln\frac{|a'c|\,|b'd|}{|a'd|\,|b'c|}\right), \tag{16}$$
where images of current sources are located at points $a'$ and $b'$ (Fig. 1b) with the distance relationships:
$$|Oa'||Oa| = R^2, \qquad |Ob'||Ob| = R^2. \tag{17}$$
One may perform calculations using (17) and get a correction proportional to $R^{-2}$ in the relation (6). However, the explicit form is too long for practical use. Instead, we present an explicit correction for the arrangements shown in Fig. 2. Minimal circles (shown in red in Fig. 2) have a unit radius and are centered on the origin of a disk with radius $R$. Correcting for finite size, eq. (6) reads
$$f(\rho, r_1, r_2, r_3, r_4, r_5) + k_2 R^{-2} + k_4 R^{-4} O(R^{-6}) = 1, \tag{18}$$
where the coefficients $k_2$ and $k_4$ can be obtained from eqs. (16) and (6), as reported in Table. 2

## 4. Experimental verification

Measurements were performed on a circular disk of high-resistivity Cr-Co-Fe-Al alloy disk with thickness of 0.05 mm and diameter 150 mm at ambient conditions without temperature stabilization. The contacts were fabricated from brass sheet of thickness 0.1 mm. Sharp spikes were formed (Fig. 5) resulting in the contacts diameter about 0.05 mm. For an arrangement of five contacts we have measured automatically all 120 possibilities of choices of four-point resistances. These measurements show consistency of results and they serve as demonstration that contacts do not move during the measurements. The apparatus and method of averaging and extracting the independent measurements is described elsewhere [6]. Average four-probe resistance was measured with automatic changes in current and voltage contacts for all permutations of *abcde*.

Using the reciprocity theorem valid for any four-probe resistances $r_{abcd} - r_{acbd} + r_{adbc} = 0$ [5], we estimated $\bar{r}_{abcd}$ and its uncertainty $\delta\bar{r}_{abcd}$ as averages

$$\bar{r}_{abcd} = \frac{1}{2}(\langle r_{abcd}\rangle + \langle r_{acbd}\rangle - \langle r_{adbc}\rangle), \tag{19}$$

$$\delta\bar{r}_{abcd} = \langle r_{abcd}\rangle - \langle r_{acbd}\rangle + \langle r_{adbc}\rangle, \tag{20}$$

where

$$\langle r_{abcd}\rangle = \frac{1}{8}(r_{abcd} + r_{badc} + r_{cdab} + r_{dcba} - r_{bacd} - r_{abdc} - r_{dcab} - r_{dcab}). \tag{21}$$

All measurements were performed with a stabilized current (100 mA, 200 mA, 300 mA, and 400 mA). It was verified that current-voltage characteristics were linear and thus we do not observe measurable effects of sample heating.

The results of two independent experiments are reported in rows 1 and 2 of Table 3. Contacts were arranged as in Fig. 2c; they were placed by hand on a 15-mm-diameter circle. The theoretical resistance values for an *abcde* contact order (Fig. 2) on an infinite plane are $(r_1, r_2, r_3, r_4, r_5) = \rho_0(+2, +1, -1, -1, +1)\ln 2/(4\pi)$. Due to imperfect contact adjustment, the agreement and reproducibility of $r_i$ measurements listed in rows 1 and 2 are poor, while the estimated sheet resistance $\rho_0$ values, from eq. (4), are close to each other. The results of another independent experiment performed for the *abdce* contact order are presented in rows 3 and 4 of Table 3. A similar experiment was performed with the contacts located on a 2.8-mm-diameter circle (rows 5 and 6, Table 3). The influence of correcting for finite size, according to eq. (17), can be observed in columns 7 and 8. All results of corrected sheet resistances in column 8 are in reasonable agreement with the results of standard van der Pauw measurements performed with contacts located on the edge of the disk: ten independent measurements yielded $\rho_0 = 25.670$ m$\Omega$.

A detailed analysis of uncertainties with eq. (7) shows that the precision of experimental sheet resistance determination should be one order of magnitude worse than the precision of the measurement of a single resistance $r_i$. Moreover, we find that the expected precision of a sheet resistance determination does not depend critically on contact arrangement and varies within a range smaller than one order of magnitude, provided that the value of the *p* parameter in eq. (13) is not too small.

## 5. Physical significance

Physical advantages of measurement methods are important. With the van der Pauw method, one measures sheet resistance on some shape with well-defined edges. It was reported that the edge may influence bulk (or planar) properties [7,8,9,10] and thus may influence the results of measurements. In this context, the presented five-point method serves as a local probe that should not be obscured by edge effects, thus overcoming the edge effect problem that can occur with the van der Pauw method.

Another advantage of the five-point method, relative to the van der Pauw method, is that with appropriate separation of contact locations, one can change the spatial scale of measurement. In particular, our method can be used as a local probe for the measurement of a planar object on a nanometer scale.

The five-point method could, potentially, have applications for use with molecular layers as a sheet resistance standard, which is important for metrology. If the layers are isotropic in sense of conducting properties, this type of sheet resistance standard would be complementary to a recently announced calculable resistance ratio [11].

Recently, a group reported two experiments to measure the sheet resistance of a spherical conducting surface: micrometer polymer spheres covered by a metallic layer [12,13]. In one experiment, they conducted four-probe measurements and relied on finite element calculations to interpret the measurements and extract the sheet resistance [12]. In another experiment, they etched a cross-shaped structure on the sphere by focused ion beam milling and measured the sheet resistance using the van der Pauw method by placing probes on the points of the cross [13]. However, because a sphere is equivalent to a plane in the sense of a Möbius transform, our five-point method can be applied directly to measure the sheet resistance of a spherical layer, without modifying the surface or relying on complex calculations. Furthermore, our method could potentially be applied to objects of micro- and nano-metric dimensions because the measurement results do not depend on knowing the specific probe positions, which are difficult to control at these scales.

The proposed five point method has clear advantage over the well-known four-probe method, in which the probes have to be precisely adjusted in line or on the vertices of square. In the proposed solution the results of the sheet resistance measurement do not dependent on the precision of the probe adjustment. Moreover, the method can be used for measurement of curved layers, for which the probe distance is not well defined.

Acknowledgments: This work was partly supported by the National Science Centre, Poland under grant OPUS no 2018/31/B/ST3/00279.

**Appendix 1**

One may consider a complex plane $\mathbb{C}$, treating $a$, $b$, $c$, $d$, and $e$ as complex numbers. From complex number analysis, it is known that potentials and currents are invariant under conformal transformations of the complex domain. The Möbius map

$$z \to f(z) = \frac{\alpha z + \beta}{\gamma z + \delta} \tag{A1}$$

is a conformal transform of the complex plane into itself. Note that the abbreviation $f$ used in (A1) should not be confused with $f$ in (2). Function $f$ in (A1) has the property of being uniquely defined by the three arguments $z_1, z_2, z_3 \in \mathbb{C}$ and their images $f_1, f_2, f_3 \in \mathbb{C}$. This property means that if one chooses three points on a complex plane ($z_1, z_2, z_3$) and three other points ($f_1, f_2, f_3$), it is possible to find an explicit form of the Möbius map, with the determinate numbers $\alpha, \beta, \gamma, \delta$, such that $f(z_i) = f_i$ for $i = 1, 2, 3$. These considerations show that for any three points $a$, $b$, and $c$, one can find a Möbius map $f$ that transforms them into 1, 0, and $-1$.

From elementary analysis, it follows that four-probe resistance expressed in complex variables

$$r_{abcd} = \frac{\rho_0}{2\pi} \ln \frac{|a - c|\,|b - d|}{|a - d|\,|b - c|}, \tag{A2}$$

where $|z|$ is the modulus of the complex number $z$ and $\rho_0$ is the sheet resistance. The symbols $a$, $b$, $c$, $d$, and $e$ have a double meaning in the paper: the label of the individual contacts (Fig. 1), and its position in the plane, expressed as a complex number in (A2). One can show by direct calculations that eq. (A2) is invariant under Möbius transformation. Thus, without loss of generality, one may consider the problem of five contacts located so that three of them have specific coordinates $a = -1$, $b = 0$, and $c = 1$, while two others are located at general positions, $d = \mu_1 + i\nu_1$ and $e = \mu_2 + i\nu_2$, where $\mu_1, \nu_1, \mu_2,$ and $\nu_2$ are real numbers. Using the abbreviations introduced in eqs. (4) and (5) we get:

$$x_1 = \frac{(1+\mu_1)^2 + v_1^2}{4(\mu_1^2 + v_1^2)}, \tag{A3}$$

$$x_2 = \frac{(1 - 2\mu_1 + \mu_1^2 + v_1^2)(\mu_2^2 + v_2^2)}{(1 - 2\mu_2 + \mu_2^2 + v_2^2)(\mu_1^2 + v_1^2)}, \tag{A4}$$

$$x_3 = \frac{4(\mu_1 - \mu_2)^2 + 4(v_1 - v_2)^2}{(1 + 2\mu_1 + \mu_1^2 + v_1^2)(1 - 2\mu_2 + \mu_2^2 + v_2^2)}, \tag{A5}$$

$$x_4 = \frac{(1 + 2\mu_2 + \mu_2^2 + v_2^2)(\mu_1^2 + v_1^2)}{(1 + 2\mu_1 + \mu_1^2 + v_1^2)(\mu_2^2 + v_2^2)}, \tag{A6}$$

$$x_5 = \frac{(1 - \mu_2)^2 + v_2^2}{4(\mu_2^2 + v_2^2)}. \tag{A7}$$

Solving eqs. (A3), (A4), (A6), and (A7) for $\mu_1, v_1, \mu_2,$ and $v_2$ unknowns, we get:

$$v_1 = \pm \frac{\sqrt{4x_1 - (1 + x_1 - x_2 x_5)^2}}{1 - 2x_1 - 2x_2 x_5}, \tag{A8}$$

$$v_2 = \pm \frac{\sqrt{4x_5 - (1 + x_5 - x_1 x_4)^2}}{1 - 2x_5 - 2x_1 x_4}, \tag{A9}$$

$$\mu_1 = \frac{x_2 x_5 - x_1}{1 - 2x_1 - 2x_2 x_5}, \tag{A10}$$

$$\mu_2 = \frac{x_5 - x_1 x_4}{1 - 2x_5 - 2x_1 x_4}. \tag{A11}$$

There are four solutions for $v_1, v_2, \mu_1,$ and $\mu_2$ for a set of five resistances and their exponentials $x_1, x_2, x_3, x_4,$ and $x_5$ (4). According to the introduced positions of contacts on the complex plane $a = -1, b = 0, c = 1, d = \mu_1 + iv_1$ and $e = \mu_2 + iv_2$, it is clear that $\mu_1$ and $\mu_2$ are uniquely determined, while there are four possibilities for the locations of $v_1$ and $v_2$. The equations (A3)–(A7) are invariant under sign changes for $v_1$ and $v_2$. Thus, only two solutions are essentially non-equivalent: $+v_1, v_2, \mu_1, \mu_2$ and $-v_1, v_2, \mu_1, \mu_2$. This fact will be important in further considerations.

Finally, introducing $v_1, v_2, \mu_1,$ and $\mu_2$ into eq. (A5), one can eliminate the relation between $x_1, x_2, x_3, x_4, x_5$ and $v_1, v_2, \mu_1, \mu_2$. As explained in the main text, two non-equivalent solutions of eqs. (A3)–(A7) produces two different relations for $x_1, x_2, x_3, x_4, x_5$:

$$0 = 1 - x_1 - x_5 - x_1 x_5 (1 - 2x_3) + x_1 x_4 (x_1 - 1) + x_2 x_5 (x_5 - 1) \tag{A12}$$
$$\pm \sqrt{4x_1 - (1 + x_1 - x_2 x_5)^2} \sqrt{4x_5 - (1 + x_5 - x_1 x_4)^2}.$$

Further evaluation of these two relations results in the highly symmetric condition

$$\sum_{i=1}^{5} (x_i - x_i x_{i+2} - x_i x_{i+1} x_{i+3} + x_i x_{i+2}^2 x_{i+4}) - x_1 x_2 x_3 x_4 x_5 = 1, \tag{A13}$$

which is a key result (6). Ranges of $i$ indexes are discussed in the text following eq. (3).

## Appendix 2

As explained in Appendix 1, without loss of generality, we consider five contacts with coordinates $a = -1$, $b = 0$, $c = 1$, $d = \mu_1 + i\nu_1$, and $e = \mu_2 + i\nu_2$, where $\mu_1, \nu_1, \mu_2,$ and $\nu_2$ are real numbers. For the already considered specific contact permutation *bacde* with $r'_1, r'_2, r'_3, r'_4, r'_5$ and $r_1, r_2, r_3, r_4, r_5$ defined according to eq. set (5), we introduce the following more convenient abbreviations: $f(\rho, r_1, r_2, r_3, r_4, r_5) - 1 \stackrel{\text{def}}{=} g(\rho, a, b, c, d, e)$.

By direct calculation, we obtain

$$\ln \frac{g(\rho, a, b, c, d, e)}{g(\rho, b, a, c, d, e)} = \frac{4\pi}{\rho} r_{abec}. \tag{A14}$$

Similarly, for three other permutations, which are in fact transpositions, we obtain:

$$\ln \frac{g(\rho, a, b, c, d, e)}{g(\rho, a, c, b, d, e)} = \frac{4\pi}{\rho} r_{bcad}, \tag{A15}$$

$$\ln \frac{g(\rho, a, b, c, d, e)}{g(\rho, a, b, d, c, e)} = \frac{4\pi}{\rho} r_{cdbe}, \tag{A16}$$

$$\ln \frac{g(\rho, a, b, c, d, e)}{g(\rho, a, b, c, e, d)} = \frac{4\pi}{\rho} r_{edac}. \tag{A17}$$

There is symmetry in eqs. (A14)-(A17): permuted elements shown in the denominator of the left side, such as $a'b'$, appear as the second pair of indices in the resistance term $r_{a'b'e'c'}$, while the $e'c'$ pair of indices in the resistance $r_{a'b'e'c'}$ neighbour the $a'b'$ pair in the $a'b'c'd'e'$ chain. The resistances in eqs. (A14)–(A17) are given explicitly by $\mu_1, \nu_1, \mu_2,$ and $\nu_2$:

$$r_{abec} = \frac{\rho_0}{4\pi} \ln \frac{(1 + \mu_2)^2 + \nu_2^2}{4(\mu_2^2 + \nu_2^2)}, \tag{A18}$$

$$r_{bcad} = \frac{\rho_0}{4\pi} \ln \frac{(1 - \mu_1)^2 + \nu_1^2}{4(\mu_1^2 + \nu_1^2)}, \tag{A19}$$

$$r_{cdbe} = \frac{\rho_0}{4\pi} \ln \frac{(\mu_1 - \mu_2)^2 + (\nu_1 - \nu_2)^2}{(1 - 2\mu_2 + \mu_2^2 + \nu_2^2)(\mu_1^2 + \nu_1^2)}, \tag{A20}$$

$$r_{edac} = \frac{\rho_0}{4\pi} \ln \frac{(1 - 2\mu_1 + \mu_1^2 + \nu_1^2)(1 + 2\mu_2 + \mu_2^2 + \nu_2^2)}{(1 + 2\mu_1 + \mu_1^2 + \nu_1^2)(1 - 2\mu_2 + \mu_2^2 + \nu_2^2)}. \tag{A21}$$

It is known from combinatorics that any $k$ element permutation is a composition of transpositions taken from the $k - 1$ element set, which serves as a basis. A change of contact order in one of the considered transposition results in a multiplication of function $g$ by an exponential function of $\rho_0/\rho$. Thus, all functions of $g$ corresponding to different contact permutations differ from each other by exponential functions of $\rho_0/\rho$, and all of them have the same set of zeros. Finally, all of the functions $f(\rho, r_1, r_2, r_3, r_4, r_5)$ defined in eq. (3), regardless of contact order, have the same set of solutions for eq. (6).

## Appendix 3

The property that if four contacts are located on a circle or on a line, then the uncertainty parameter $u$ in eq. (7) does not depend on the position of the fifth contact can be proven with the abovementioned arguments related to the Möbius transform. Without loss of generality, let us assume that the locations of contacts $a$, $b$, and $c$ on a complex plane are $a = -1$, $b = 0$, and $c = 1$, while contacts $d$ and $e$ are located at the general positions $d = \mu$ and $e = \mu_2 + i\nu_2$, where $\mu, \mu_2,$ and $\nu_2$ are real numbers. Thus $a, b, c, d$ are collinear. Direct calculation of $u$ with eq. (7) shows that the result depends only on $\mu$ and not on $\mu_2$ or $\nu_2$:

$$u = \frac{2\pi\sqrt{1 + \mu(9\mu - 2)}}{(1-\mu)\ln\frac{1+\mu}{1-\mu} + 2\mu\ln\frac{1+\mu}{2\mu}}, 0 < \mu < 1. \tag{A22}$$

A minimum value of $u$ (A9) is achieved for $\mu = \mu_0 \approx 0.273$, where $\mu_0$ is the root of the equation:

$$\mu\ln\frac{2\mu(1-\mu)^4}{(1+\mu)^5} + \ln\frac{1+\mu}{2\mu} = 0, 0 < \mu < 1, \tag{A23}$$

and the minimal value is $u \approx 7.66$.

**Figures**

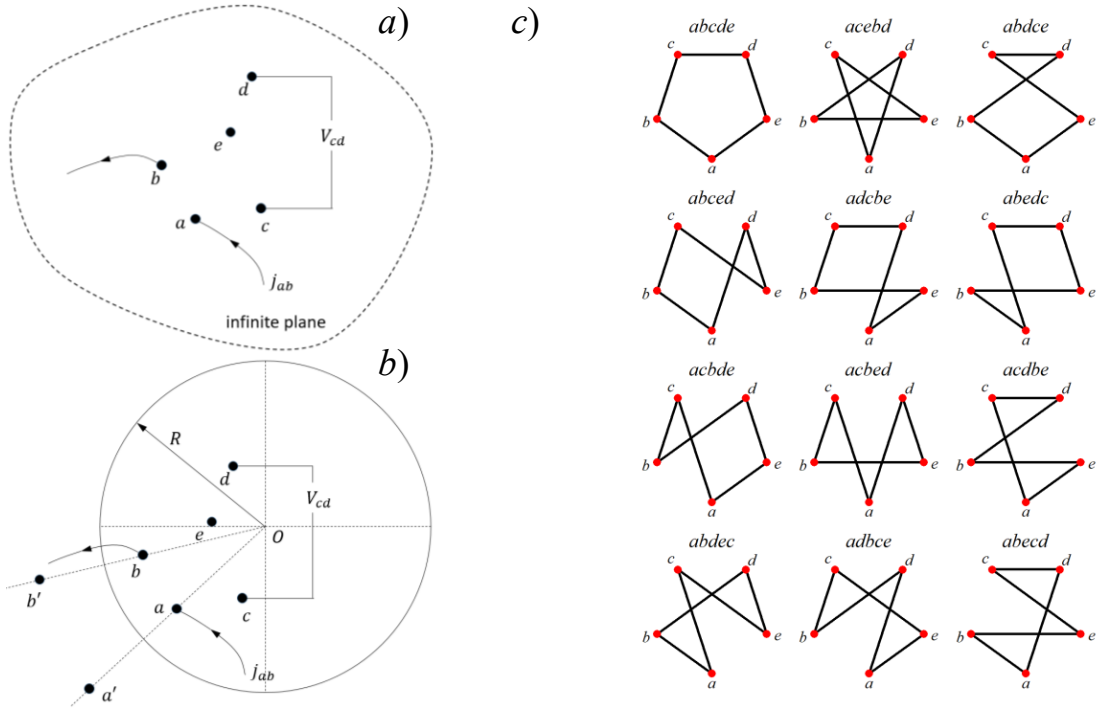

**Fig. 1** *a*) Five point contacts at positions *abcde* on *a*) an infinite plane and *b*) a disk with radius $R$. Active contacts used in the measurement of resistance $r_{abcd}$ are shown. Positions of image current sources are shown at points *a'* and *b'*. *c*) The physical meaning of twelve non-equivalent

permutations of contacts *abcde*, shown by black lines connecting the vertices and forming a loop (see text). For esthetic simplicity, we elect to use the line pattern that forms a regular pentagon.

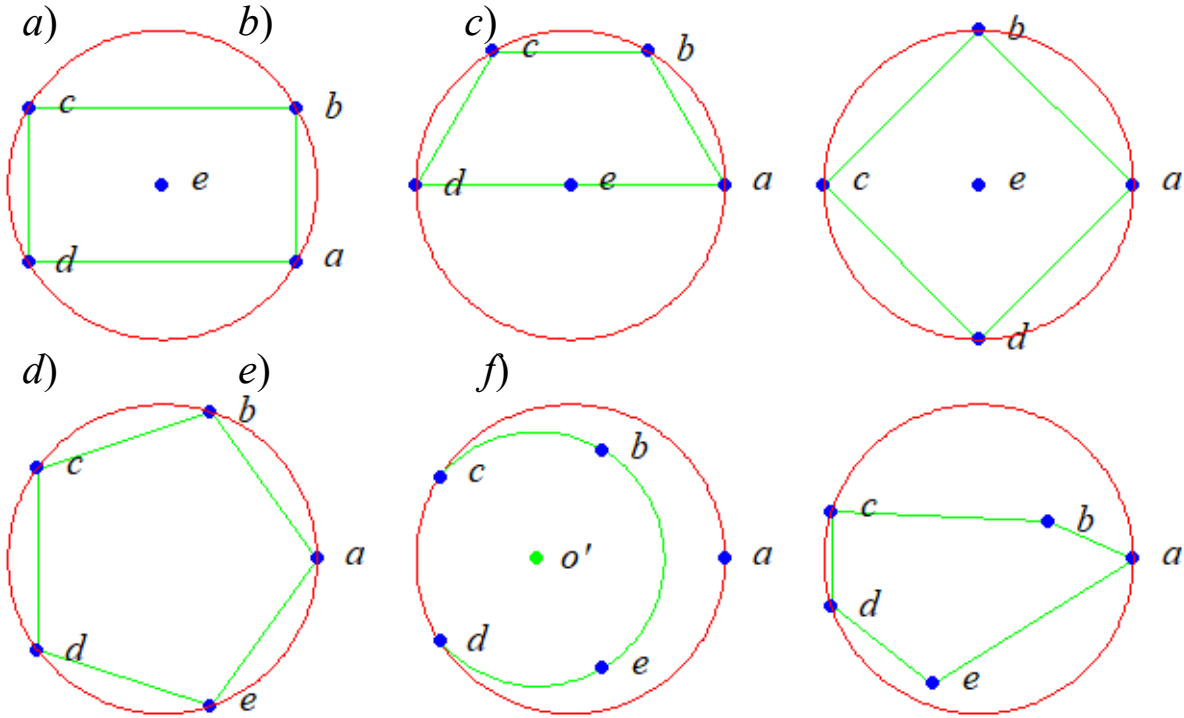

**Fig. 2.** Selected contact arrangements (blue dots) in an infinite plane. (*a–f*) Arrangements with the following specified properties: (a) triangles *abe* and *cde* are regular; (b) triangles *abe*, *bce*, and *cde* are regular; (c) *abcd* is a square; (d) *abcde* is regular pentagon; (e) contacts form an elongated pentagon with a symmetry axis, with angles *ao'e* = *co'a* = *do'c* = 80.55°, and *o'a* = 1.479*o'b*; and (*f*) coordinates {*a, b, c, d, e*} are {(1, 0), (0.446, 0.244), (−0.953, 0303), (−.953, −0.302), (−0.300, −0.809)}, respectively. Green shapes are shown to guide the eye. Minimal encompassing circles with a radius equal to 1 are shown in red; the origins of each red circle correspond to point *O* in eq. (17).

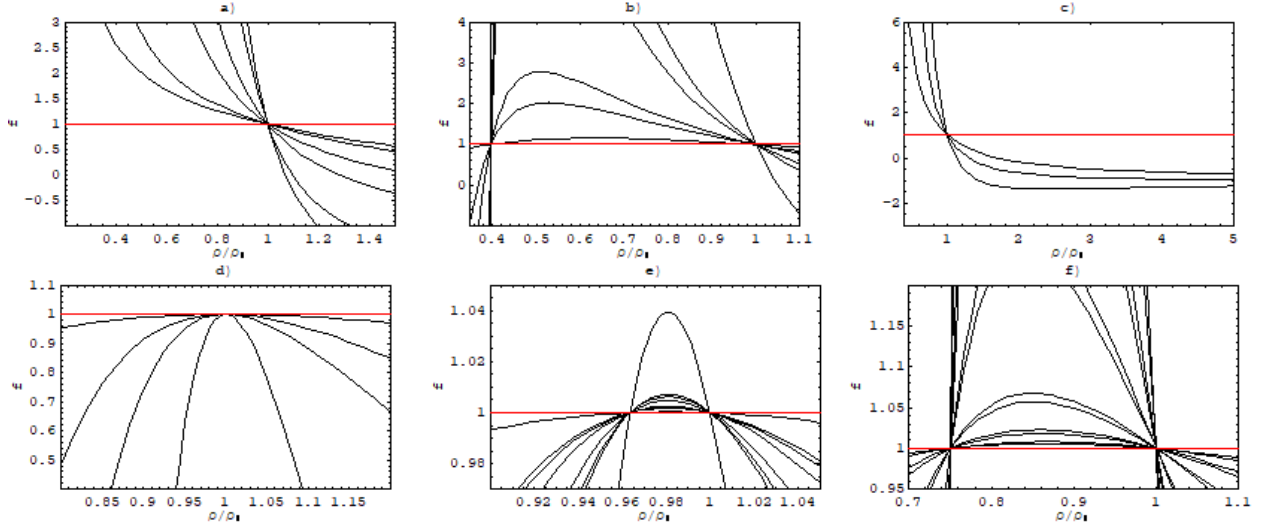

**Fig. 3.** Values of $f(3)$ plotted as a function of $\rho$ for arrangements of contacts *abcde* shown in Fig. 2 and for the nonequivalent permutations of *abcde* listed in Table 1. Values of $r_i$ (5) in eq. (6) correspond to the resistances of contacts located on a plane with sheet resistance $\rho_0$. It can be seen that some curves corresponding to nonequivalent permutations overlap. The number of non-overlapping curves is 6, 6, 3, 4, 8, 12 for *a)*, *b)*, *c)*, *d)*, *e)*, and *f)*, respectively. All curves cross the red line at $\rho = \rho_0$. For the contact arrangements shown in *b)*, *e)*, and *f)*, there are nonphysical solutions of eq. (6) for $\rho < \rho_0$.

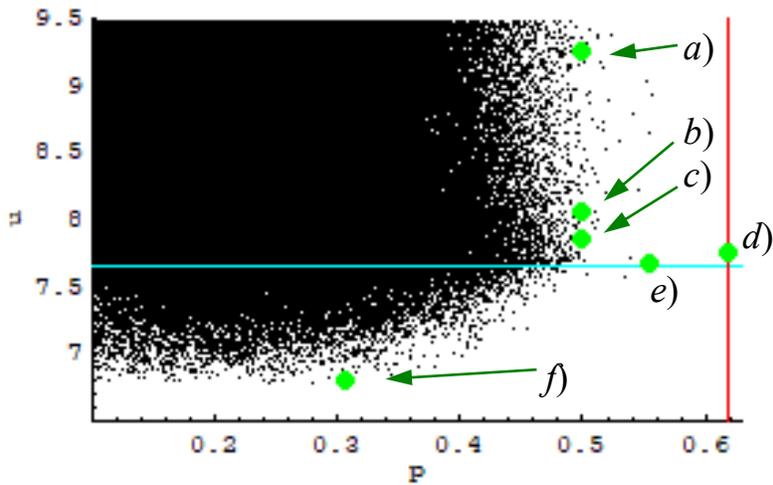

**Fig. 4.** Scatter plot of the uncertainty parameters $u$ in eq. (7) and the packing ratios $p$ in eq. (13) obtained from simulations for five randomly located contacts (black points). Green points correspond to arrangements with the corresponding letter labels in Fig. 3 and Table 1. The vertical red line shows the maximum packing ratio for five in-plane points (regular pentagon). The horizontal cyan line shows the minimum value of the uncertainty parameter $u$ for four points located in a circle.

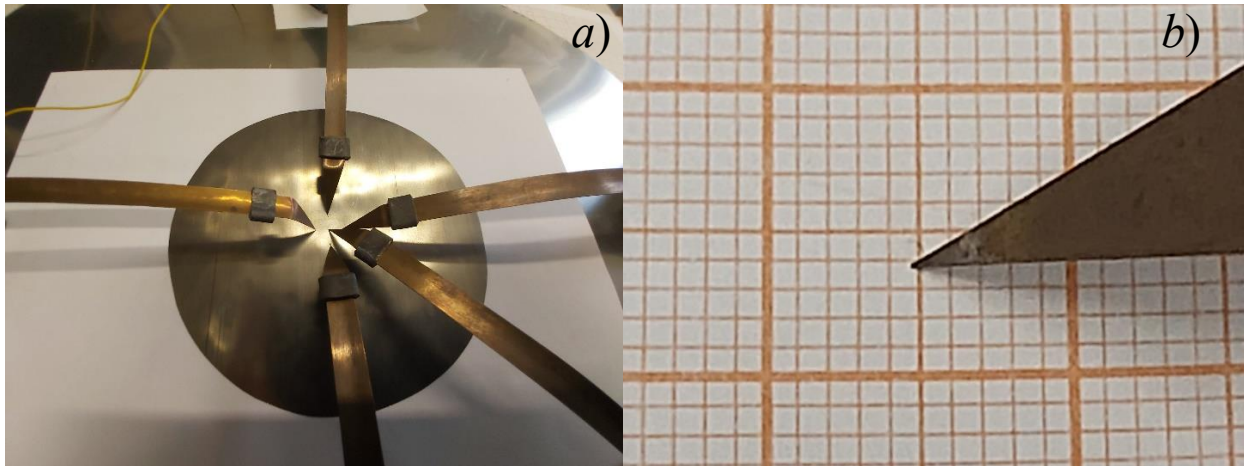

**Fig. 5.** Experimental setup for five point measurements, *a)* five contacts on metallic disc, *b)* zoom of the single contact.

**Table 1.** Uncertainty parameter $u$ (eq. 7) values for nonequivalent permutations of contacts arranged as in Fig. 2, with those permutations yielding best $u$ value highlighted in yellow.

|  | **Configuration in Fig. 2** | | | | | |
|---|---|---|---|---|---|---|
|  | *a)* | *b)* | *c)* | *d)* | *e)* | *f)* |
| Packing ratio $p$ | 0.5 | 0.5 | 0.5 | 0.618 | 0.554 | 0.306 |
|  | | | **Column** | | | |
| 1 | 2 | 3 | 4 | 5 | 6 | 7 |
| Contact order | | | $u$ values | | | |
| *abcde* | 9.2645 | 8.0598 | 7.8503 | 7.7553 | 7.6640 | 7.9960 |
| *abced* | 12.1759 | 9.8712 | 7.8503 | 10.7385 | 10.7607 | 12.2412 |
| *abdce* | 11.8512 | 10.9131 | 11.1020 | 14.5819 | 14.0250 | 12.2788 |
| *abdec* | 16.0465 | 14.3426 | 13.5971 | 13.8405 | 13.5932 | 13.1827 |
| *abecd* | 9.2645 | 8.0598 | 7.8503 | 12.5484 | 11.8171 | 11.0600 |
| *abedc* | 11.8512 | 10.9131 | 11.1020 | 9.0121 | 8.4061 | 7.7088 |
| *acbde* | 17.8861 | 15.4334 | 13.5971 | 10.7385 | 10.7607 | 9.8877 |
| *acbed* | 16.2879 | 13.5667 | 11.1020 | 7.7553 | 7.6640 | 8.3260 |
| *acdbe* | 16.0465 | 14.3426 | 13.5971 | 12.5484 | 11.8171 | 8.3811 |
| *acebd* | 17.8861 | 15.4334 | 13.5971 | 14.5819 | 14.0250 | 12.4962 |
| *adbce* | 16.2879 | 13.5667 | 11.1020 | 13.8405 | 13.5932 | 12.6754 |
| *adcbe* | 12.1759 | 9.8712 | 7.8503 | 9.0121 | 8.4061 | 6.8050 |

**Table 2.** Correction for finite size (eq. 14) for nonequivalent permutations of contact arrangements shown in Fig. 2.

| Configuration in Fig. 3 | *a)* | | *b)* | | *c)* | | *d)* | | *e)* | | *f)* | |
|---|---|---|---|---|---|---|---|---|---|---|---|---|
|  | $k_2$ | $k_4$ | $k_2$ | $k_4$ | $k_2$ | $k_4$ | $k_2$ | $k_4$ | $k_2$ | $k_4$ | $k_2$ | $k_4$ |
| *abcde* | 3 | −3 | 1 | −3 | 8 | 0 | 0 | −5.04 | 0.15 | −3.63 | 2.54 | -8.16 |
| *abced* | 9 | 9 | 4 | 0 | 8 | 0 | 0 | −34.5 | 1.14 | -22.9 | 2.92 | -8.53 |
| *abdce* | 4 | 0 | 3 | −3 | 16 | 32 | 0 | −34.5 | 0.82 | -17.2 | 15.59 | -20.78 |
| *abdec* | 16 | 48 | 9 | 9 | 32 | 128 | 0 | −90.5 | 2.66 | -49.2 | 0.27 | -1.77 |
| *abecd* | 3 | −3 | 4 | 0 | 8 | 0 | 0 | −34.5 | 2.02 | -38.6 | 6.40 | -14.81 |
| *abedc* | 4 | 0 | 3 | −3 | 16 | 32 | 0 | −34.5 | 0.87 | -17.9 | 0.097 | -0.75 |
| *acbde* | 48 | 240 | 9 | −3 | 32 | 128 | 0 | −34.5 | 1.14 | -22.9 | 0.34 | -1.85 |
| *acbed* | 36 | 144 | 12 | 24 | 16 | 32 | 0 | −90.5 | 2.82 | -51.0 | 0.14 | -0.89 |
| *acdbe* | 16 | 48 | 9 | 9 | 32 | 128 | 0 | −90.5 | 2.01 | -38.6 | 0.74 | -3.60 |
| *acebd* | 48 | 240 | 36 | 144 | 32 | 128 | 0 | −617. | 15.3 | -230. | 0.84 | -3.89 |
| *adbce* | 36 | 144 | 12 | 24 | 16 | 32 | 0 | −90.5 | 2.66 | -49.2 | 22.06 | -1.48 |
| *adcbe* | 9 | 9 | 4 | 0 | 8 | 9 | 0 | −34.5 | 0.87 | -17.9 | 7.88 | -10.57 |

**Table 3.** Resistance measurement results for configurations shown in Fig. 2*c* with an inner circle diameter $\phi$; results of analysis related to eqs. (6), (18), and (7) are shown in columns 7, 8, and 9, respectively, while an approximate diameter of the inner circle for contact location is shown in column 10.

| Points | $r_1$ [mΩ] | $r_2$ [mΩ] | $r_3$ [mΩ] | $r_4$ [mΩ] | $r_5$ [mΩ] | $\rho$ [mΩ] | $\rho$ [mΩ] | $\delta\rho$ [mΩ] | $\phi$ [mm] |
|---|---|---|---|---|---|---|---|---|---|
| 1 | 2 | 3 | 4 | 5 | 6 | 7 | 8 | 9 | 10 |
| *abcde* | +2.860(1) | +1.830(1) | −1.491(2) | −1.6730(7) | +1.094(1) | 26.25 | 24.92 | 0.01 | 15 |
| *abcde* | +3.053(2) | +0.877(2) | −1.181(1) | −1.222(1) | +1.855(3) | 26.19 | 24.58 | 0.01 | 15 |
| *abdce* | −3.0309(7) | −0.9874(13) | +1.145(1) | +1.8502(13) | +0.708(1) | 26.19 | 25.84 | 0.01 | 15 |
| *abdce* | −1.851(1) | −1.886(1) | +0.863(2) | +3.0298(7) | +1.8874(6) | 26.18 | 25.57 | 0.01 | 15 |
| *abcde* | +3.682(1) | +1.1031(8) | −0.184(6) | −0.256(2) | +1.037(2) | 25.74 | 25.68 | 0.02 | 2.8 |
| *abcde* | +3.053(2) | +0.877(2) | −1.182(1) | −1.222(1) | +1.855(3) | 26.19 | 26.16 | 0.01 | 2.8 |

# References

[1] M. Reveil, V. C. Sorg, E. R. Cheng, T. Ezzyat, P. Clancy and M. O. Thompson, Finite element and analytical solutions for van der Pauw and four-point probe correction factors when multiple non-ideal measurement conditions coexist, Rev. Sci. Instr., 88 (2017) 094704

[2] Z.-H. Sun, J. Zhou, X.-J. Xia and D.-M. Zhou, Two-dimensional electrostatic model for the Van der Pauw method, Phys. Lett. A, 381 (2017) 2144

[3] G. González-Díaz, D. Pastor, E. García-Hemme, D. Montero, R. García-Hernansanz, J. Olea, A. del Prado, E. San Andrés and I. Mártil, A robust method to determine the contact resistance using the van der Pauw set up, Measurement, 98 (2017) 151

[4] I. Miccoli, F. Edler, H. Pfnür and C. Tegenkamp, J. Phys.: Condens. Matter, 27 (2015) 223201

[5] L. J. van der Pauw, A Method of Measuring Specific Resistivity and Hall Effect of Discs of Arbitrary Shape, Phil. Res. Reports, 13 (1958) 1

[6] K. Szymański and P. Zaleski, Precise Measurement of Inhomogeneity of 2-D System by Six-Point Method, IEEE Trans Instr. Meas., 66 (2017) 1243

[7] K. Wang, C. Wu, Y. Jiang, D. Yang, K. Wang and S. Priya, Distinct conducting layer edge states in two-dimensional (2D) halide perovskite, Sci. Adv., 5 (2019) 1

[8] K. Sasaki, S. Saito, R. Saito, Stabilization mechanism of edge states in graphene, Appl. Phys. Lett., 88 (2006) 113110

[9] Anup Pramanik, Bikash Mandal, Sunandan Sarkar, Pranab Sarkar, Effect of edge states on the transport properties of pentacene–graphene nanojunctions, Chem. Phys. Lett., 597 (2014) 1

[10] A. Ostovan, Z. Mahdavifar, M. Bamdad, Length–dependence of conductance in benzothiadiazolemolecular wires between graphene nanoribbon electrodes: Effect of conformational changes, J. Mol. Liq., 269 (2018) 639

[11] K. M. Yu, D. G. Jarrett, A. D. Koffman, A. F. Rigosi, S. U. Payagala, K.-S. Ryu, J.-H. Kang and S.-H. Lee, Using a Natural Ratio to Compare DC and AC Resistances, IEEE Tran. Instr. Meas., 2020

[12] M. Bazilchuk, O. M. Evenstad, Z. Zhang, H. Kristiansen and J. He, Resistance Analysis of Spherical Metal Thin Films Combining Van Der Pauw and Electromechanical Nanoindentation Methods, J. Electr. Mat., 47 (2018) 6378

[13] S. R. Pettersen, A. E. Stokkeland, H. Kristiansen, J. Njagi, K. Redford, D. V. Goia, Z. Zhang, and J. He, Electrical four-point probing of spherical metallic thin films coated onto micron sized polymer particles, Appl. Phys. Lett., 109 (2016) 043103